\documentclass[10pt]{revtex4}
\usepackage{graphicx}

\begin{document}

\title{Observing The Observer}
\author{Vlatko Vedral}
\affiliation{Clarendon Laboratory, University of Oxford, Parks Road, Oxford OX1 3PU, United Kingdom\\Centre for Quantum Technologies, National University of Singapore, 3 Science Drive 2, Singapore 117543\\
Department of Physics, National University of Singapore, 2 Science Drive 3, Singapore 117542\\
Center for Quantum Information, Institute for Interdisciplinary
Information Sciences, Tsinghua University, Beijing, 100084, China}

\begin{abstract}
I present a simple variant of the Schr\"odinger cat meets Wigner's friend thought experiment. If you are shocked by it, you have not understood quantum physics (no words are missing from
this sentence). 
\end{abstract}

\maketitle

The idea that quantum mechanics applies to everything in the universe, even to us humans, can lead to some interesting conclusions \cite{Vedral}.

Consider Deutsch's variant \cite{Deutsch} of the iconic Schr\"odinger cat thought experiment that builds on Wigner's ideas \cite{Wigner}. I will try to use as interpretation-free language as possible and let you draw your own conclusions. I've personally witnessed all sorts of reactions from physicists upon hearing this. I've seen a conversion from Copenhagen to Many-Worlds (which is the quantum equivalent of a religious conversion from, say, Christian Orthodoxy to Buddhism). At the other extreme, I've heard colleagues say: ``so what's the big deal with this stuff?" (The latter reply, in my experience, typically comes either from those who truly understand quantum physics or those who completely missed the point). 

Suppose that a very able experimental physicist, Alice, puts her friend Bob inside a room with a cat, a radioactive atom and cat poison that gets released if the atom decays. The point of having a human there is that we can communicate with him. (Getting answers from cats is not that easy - believe me, I've tried.) As far as Alice is concerned, the atom enters into a state of being both decayed and not decayed, so that the cat is both dead and alive (that's where Schr\"odinger stops). 

Bob, however, can directly observe the cat and sees it as one or the other. This is something we know from everyday experience: we never see dead and alive cats. To confirm this, Alice slips a piece of paper under the door asking Bob whether the cat is in a definite state. He answers, ``Yes, I see a definite state of the cat".

At this point, mathematically speaking, the state of the system has changed from the initial state
\begin{equation}
|\Psi_i\rangle = |\textnormal{no-decay}> |\textnormal{poison in the bottle}> |\textnormal{cat alive}> |\textnormal{Bob sees alive cat}> |\textnormal{blank piece of paper}>
\end{equation}
to the state (from Alice's global perspective):
\begin{eqnarray}
|\Psi_{1/2}\rangle = & & (|\textnormal{decay}> |\textnormal{poison released}> |\textnormal{cat dead}> |\textnormal{Bob sees dead cat}> +\nonumber \\
& & |\textnormal{no-decay}> |\textnormal{poison still in the bottle}> |\textnormal{cat alive}> |\textnormal{Bob sees alive cat}>) \otimes \nonumber \\
& & |\textnormal{paper says: ”yes, I see a definite state of the cat”}>
\end{eqnarray}

I am assuming that, because Alice's laboratory is isolated, every transformation leading up to this state is unitary. This includes the decay, the poison release, the killing of the cat and Bob's observation - Alice has a perfect quantum coherent control of the experiment. 

Note that Alice does not ask whether the cat is dead or alive because for her that would force the outcome or, as some physicists might say, “collapse” the state (this is exactly what happens in Wigner's version, where he communicates the state to a friend, who communicates to another friend and so on...). She is content observing that Bob sees the cat either alive or dead and does not ask which it is. Because Alice avoided collapsing the state, quantum theory holds that slipping the paper under the door was a reversible act. She can undo all the steps she took since each of them is just a unitary transformation. In other words, the paper does not get entangled to the rest of the laboratory. 

When Alice reverses the evolution, if the cat was dead, it would now be alive, the poison would be in the bottle, the particle would not have decayed and Bob would have no memory of ever seeing a dead cat. If the cat was alive, it would also come back to the same state (everything, in other words, comes back to the starting state where the atom has not decayed, the poison is in the bottle, the cat is alive and Bob sees alive cat and has no memory of the experiment he was subjected to). 

And yet one trace remains: the piece of paper saying ``yes, I see a definite state of the cat". Alice can undo Bob's observation in a way that does not also undo the writing on the paper. The paper remains as proof that Bob had observed the cat as definitely alive or dead. (Note that I am still remaining interpretation neutral. A Many Worlds supporter would say that there are two copies of Bob, one that observes a dead cat and one that sees alive cat; a Copenhagen or Quantum Bayesian supporter could say that relative to one state of Bob the cat is dead, while, relative to the other, it is alive - either way, supporters of any interpretation ought to make the same predictions in this experiment).

That leads to a startling conclusion for someone who believes that measurements have definite irreversible outcomes. Alice was able to reverse the observation because, as far as she was concerned, she avoided collapsing the state; to her, Bob was in just as indeterminate a state as the cat. But the friend inside the room thought the state did collapse. That person did see a definite outcome; the paper is proof of it. In this way, the experiment demonstrates two seemingly contradictory principles. Alice thinks that quantum mechanics applies to macroscopic objects: not just cats but also Bobs can be in quantum limbo. Bob, halfway through the experiment and before Alice's reversal, thinks that cats are only either dead or alive.

I deliberately avoided interpretational jargon because this experiment cannot distinguish between different interpretations (nothing can, that's why they are called interpretations). What it does is tell the difference between quantum physics being valid at macroscopic scales and there being a genuine collapse due to observation. If Bob genuinely collapsed the quantum state inside by seeing one outcome or the other (definitely ``dead" or definitely ``alive"), then the reversal would not be possible with unit probability. Namely, both outcomes would occur at the end of the experiment, the atom could also be found decayed and the cat dead (with one half probability).  Therefore repeating this experiment a few times would tell us if the observation leads to a definitive collapse or not. 

Doing such an experiment with an entire human being would be daunting, but physicists can accomplish much the same with simpler systems. We can take a photon and bounce it off a mirror. If the photon is reflected, the mirror recoils, but if the photon is transmitted, the mirror stays still. The photon plays the role of the decaying atom; it can exist simultaneously in more than one state. The mirror, made up of many atoms, acts as the cat and as Bob. Whether it recoils or not is analogous to whether the cat lives or dies and is seen to live or die by Bob. The process can be reversed by reflecting the photon back at the mirror. If the photon always comes out the way it came in, we confirmed that it was in a superposition after the mirror and before the reversal. Otherwise, there was a collapse somewhere along the way (needless to say, there is no collapse if things are done properly in actual photonic experiments of this kind). 

We can do similar experiments with (collections of) atoms and molecules, where we entangle them and subsequently disentangle them. Again, no collapse recorded. 

In developing this devious thought experiment, Wigner and Deutsch followed in the footsteps of Schr\"odinger, Einstein, Bell and other theorists who argued that physicists have yet to grasp quantum mechanics in a deep way. For decades most physicists scarcely cared because the foundational issues had no effect on practical applications of the theory. But now that we can perform these experiments for real, the task of exploring the full extent of quantum mechanics has become all the more urgent. 

{\em Disclosure of potential CoI}: to me personally, the validity of quantum physics at the macroscopic level naturally suggests the Many Worlds picture. The Many Worlds interpretation is not without problems, such as the (lack of a convincing) derivation of the Born rule and (in my opinion much less of an issue, but still...) the basis problem. However, other interpretations have much the same concerns (which they avoid by calling the Born rule a postulate and saying that the coupling to the environment selects the basis, therefore eliminate the need for explanation). The simple fact demonstrated here, however, namely that from one perspective (Bob's above) we can have definite outcomes, while, at the same time, from a higher perspective (Alice's above) everything remains in a quantum coherent state, seems to me to be spiritually leaning towards Many Worlds. 

And now, I will stop, shut up and (continue to) experiment.

\textit{Acknowledgments}: The author acknowledges funding from the National Research Foundation
(Singapore), the Ministry of Education (Singapore), the Engineering and Physical Sciences Research Council (UK), the Templeton Foundation, the Leverhulme Trust, the Oxford Martin School, and Wolfson
College, University of Oxford. This research is also supported by the National Research Foundation, Prime Minister’s Office, 
Singapore under its Competitive Research Programme (CRP Award No. NRF- CRP14-2014-02) and administered by 
Centre for Quantum Technologies, National University of Singapore.


\begin{thebibliography}{99}
%
\bibitem{Vedral} A much less formal exposition of the same was presented in V. Vedral, ``Living in a Quantum World", Scientific American, June 2011. 
%
\bibitem{Deutsch} D. Deutsch, ``Three experimental implications of the Everett interpretation" in R. Penrose and C.J. Isham (eds.), Quantum Concepts of Space and Time, Oxford: The Clarendon Press, pp. 204–214 (1986).
%
\bibitem{Wigner} E.P. Wigner, "Remarks on the mind-body question", in: I.J. Good, ``The Scientist Speculates", London, Heinemann (1961).


\end{thebibliography}
\end{document}